# The Principle of Synergy and Isomorphic Units, a revised version

Edgar Paternina

**Electrical engineer.** With experience in the area of real time control centers with State Estimator included, an algorithm that validates the complex equations of a power system.

**Contact: epaterni@epm.net.co**

___________________________________________________________________

**Abstract:** A solution to the part and whole problem is presented in this paper by using a complex mathematical representation that permits to define the Holon concept as a unit that remains itself in spite of complex operations such as integration and derivation. This can be done because of the remarkable isomorphic property of Euler Relation. We can then define a domain independent of the observer and the object where the object is embedded. We will then be able to have a Quantum Mechanics solution without the "observer drawback", as Karl R. Popper tried to find all his life but from the philosophical point of view and which was Einstein main concern about QM. A unit that has always similar or identical structure or form, despite even complex operations such as integration and derivation, is the ideal unit for the new sciences of complexity or just the systems sciences where structure or form, wholeness, organization, and complexity are main requirements. A table for validating the results obtained is presented in case of the pendulum formula.

___________________________________________________________________

## 1. Introduction: Three basic attributes of reality

In his <u>General System Theory</u>**,** Ludwig von Bertalanffy wrote:

*Reality, in the modern conception, appears as a tremendous hierarchical order of organized entities, leading, in a superposition of many levels, from the physical and chemical to biological and sociological systems. Unity of Science is granted, not by a utopian reduction of all sciences to physics and chemistry, but by the structural uniformity of the different levels of reality.*

That structural uniformity or isomorphism of the different levels of reality is the main concern of this paper, and its main aim will be to present a new way of "seeing" reality by means of some *isomorphic units, or co-variant units*, so to speak, units in which the form is one important attribute as well. A basic recurrent design or pattern that can be used to interpret and explain those problems where dynamic interactions or an organized complexity appear.

The problem of form appeared in classical physics, but precisely in those fields where the ***field*** concept was unavoidable. The magnetic field problem, whose existence can even be felt by putting two permanent magnets near by, is really one of those problems of nature that after all were hiding a great mystery. With the magnet we can not only present theoretical examples of those three basic fundamental attributes that are the basic to an isomorphic unit, but the magnetic field to be well represented, from the mathematical point of view, we must also use a complex number mathematical symbolism. Those three basic attributes are: wholeness, oneness and openness.

The wholeness attribute can be seen easily in the case of a magnet, where each magnet split from another magnet is precisely a whole new magnet. *This new "part" that came from an old whole is a whole too*. To obtain a whole from another whole is like to obtain a son from a father, or an object from the instantiation of a class and in this same sense Ken Wilber[8] wrote *To be a part of a larger whole means that the whole supplies a principle(or some sort of glue) not found in the isolated parts alone, and this principle allows the parts to join, to link together, to have something in common, to be connected, in ways that they simply could not be on their own...When it is said that "the whole is greater than the sum of its parts," the "greater" means "hierarchy"...This is why "hierarchy" and "wholeness" are often uttered in the same sentence*

Associated with this wholeness attribute is that binary or dual aspect of reality, where we have always two "opposites" or more appropriately, complementary or polar entities within a comprehensive whole, or just a unit that transcends duality, the oneness attribute, that can be found physically in a magnet. The within and the without [4] some sort of dynamic structure embedded in a unit that cannot be split in its two components. Structures that perform well in a changing environment must be capable to reflect the without, they must have, as it were, a *storing capacity* to reflect that without.

The magnet has also another remarkable attribute associated with form, with the environment or interface or some sort of a medium to separate an internal milieu from an external environment[13], as it were, a field concept, or the openness attribute. At this point is important to recall that these three attributes cannot be considered on their own. They are linked together by some sort of glue. This is the main characteristic of this new way of "seeing" reality in which we must always have in mind that gluing principle, we will name the Principle of Synergy. At this point of reasoning we are taken to think in open systems, as those *systems capable of exchanging with the environment*

## 2. Thirdness as an alternative way of codification

For representing mathematically such kind of problems, it is necessary to use a language that permits to define a unit embedded in a comprehensive whole, environment or dynamic structure, and which can include also a radical duality or polarity, which can be generated, precisely because that inherent coincidentia oppositorum or that tension, that manifests as a field that makes by definition that unit an open system: *the self-preservation attribute*.

The main aim of this semiotic codification[16] is to integrate in a comprehensive whole the radical duality of the universe that can be codified as

- (the original, (the potential/the actual))

- (Being or reality, (mind or consciousness/form)

- (Form, (time/space))

- (wholeness, (Oneness/Openness))

where:
- with *openness* we associate the field concept,

- with *oneness* the dual nature properly speaking not in the monadic sense**,**

- and with *wholeness* the nondual or qualitative nature of reality, in the thirdness sense.

It is important to notice that we have two fundamental entities:

- one that stands on its own and

- another one dual embedded in another parenthesis and separated by a symbol "/" of the or type in the sense that the one excludes the other in its manifestation The symbol "," is for differentiating two different and fundamental orders of reality, and the symbol ( ), is for integrating them both in a comprehensive whole. It is the relation between these two components that permits us to redefine the uncertainty principle in a new context as we will see later.

A Holon has embedded a within and a without, as it were, an analytical and a synthetic capacity, or a partness and a wholeness attribute.

The openness attribute is then related with a capacity to generate a field that is concomitant with those entities we can named holonic -to use a term coined by Ken Wilber- such as magnetic entities, electrons, linguistic signs, cells, life, mind and beings in general.

## 3. Complex number and thirdness

Complex numbers have the capacity not only to represent that binary or duality aspect of reality or the chance to have two polarities included in one unit, but also the nondual or wholeness attribute, we have related with that capacity to generate a new whole, *the self-replication attribute*.

Complex numbers were born when trying to solve the simple algebraic equation

$$x^2 + 1 = 0$$

$$x^2 = -1$$

$$x = Sqr(-1) = J$$

where as a solution we have the square root of a negative one or the radical unit **J**.

From the point of view of "real" number perspective it does not exist a solution for this simple problem and as so it was necessary to make a paradigm extension or paradigm shift, and to define a new type of more general numbers to solve the problem. The solution J was named "imaginary" by Descartes for the first time and since then, complex numbers remained as some sort of a strange mathematical tool. It was Leonard Euler in 1745, the one, that finally found a mathematical symbol for representing that new entity that included both kinds of numbers:
- those named real and which we will term nondual for reasons we will see later, and
- those named "imaginary", we will name dual on the other hand for apparent reasons too.

Why was it written by David Berlinski in "Tour of the Calculus"(1196, Heinemann) that "the area in thought that the calculus made possible is coming to an end" and that "it is a style that has shaped the physical but not the biological science"? A reason is that area in thought of classical mathematics was definitely restricted to the without of things, to the surface, to the reduced part. And as so it could only be codified from a monadic world perspective and in a way that the within necessarily had to be reduced to the without of things[16 ], the quantitative to the qualitative. In such a monadic world we have:
- (within/without)... reduced to... (without)
- (qualitative/quantitative)... reduced to... (quantitative)
- (whole/part)... reduced to... (part)

- (potential/actual)... reduced to... (actual)

Being the main advantage of this mathematical methodology or language that with it, we can work with close systems, reversible, universal, deterministic and atomistic [16]

Problems of growth has been associated from the point of view of a mathematical representation with the number epsilon or Euler number since a long time ago. And it was by studying infinite series, that Euler found that entity, that not only could represent those two kinds of numbers, nondual and dual, but also it included those cyclical waveforms, sine and cosine, that occur so frequently in nature wherever we have cyclical phenomena. But the most important, the most cogent argument to use this kind of new numbers, that have been used by Electrical Engineering since Oliver Heaviside and Steinmetz introduced them at the end of the 19th century to solve alternating electrical current circuits, is precisely, its inherent isomorphic property, that permits them to make, as it were, co-variant representations, or to make simpler complex operations.

Evidently such a useful mathematical entity, was adequate to represent dynamic realities, and it was used for the first time for representing electromagnetic fields at the end of the 19th century, and after that, vectors were born in physics, but then they are taught, in general, without making any references to this complex number origin. Up to that moment complex numbers had not been used in practical cases, and as so it justifies why the complex plane was delayed 100 years from its real birth at the end of the 18th century.

A unit that has always similar or identical structure or form, despite even complex operations such as integration and derivation, is the ideal unit for the new sciences of complexity or just the systems sciences where structure or form, wholeness, organization, and complexity are main requirements. Another important point is that it can also be used to define then a Basic Unit System concept in which uncertainty is included, and as so open systems.

Classical physics had as it main aim to resolve natural phenomena into a play of elementary units, as it were, to resolve those phenomena into their parts, isolated parts, I mean, so the concept of particle was always the starting point of the whole framework. But that part or particle needed to be considered as an isolated entity, that is, as a closed system, in which there were no interactions at all with the environment[2]. This ideal model to represent reality was so restrictive that it definitely failed, the way we all know.

An adequate framework for representing reality, the whole reality, must be complex, in the sense, that it must not only include, the dual-logical-nature of reality, but also it must include, that another aspect related with form, wholeness and oneness.

The form, the structure must be co-variant[5], that is, it must remain the same in spite of a progressive modification of that same structure. In this way adaptation or changeness as a fundamental process can give us persistent properties for that structure or just despite complex operations such as integration and derivation done upon that structure. There is another more important aspect to recall, and it is the need not to reduce uncertainty, in certain cases, to be able just to manage it. And here we come across with the fundamental problem of open and closed systems. When we have an open system, in general, that uncertainty is an essential part of the problem, of its openness attribute, and in this sense that uncertainty cannot be reduced unless we close the system, so that we determine its state completely defining then ideal objects of study, that can fail in real cases. The interactions of a closed system are reduced almost to zero and then the system becomes a static system and not precisely a steady state system. Open systems and that second law are in some sense "incompatible" if we establish a hierarchical framework in which, that second law is just a special case of the behavior of an open system.

## 4. Euler Relation and Its Isomorphic Properties

Up to this moment we have been seeing the emergence of a new concept of unit that includes in its mathematical representation the dual and nondual nature of reality, i.e., a relationship between the part and the whole or just a unit that is a whole and a part at the same time or a Holon as Ken Wilber named it [8], represented succinctly as:

(whole, part)

What we aim at this moment is to show that Euler Relation, is precisely the mathematical symbol necessary to make an adequate mathematical representation of the Holon concept by taking J as the symbol for differentiation. The whole/part entity we have named a Basic Unit System can be codified then as:

$$e^{J(\emptyset)} = \cos(\emptyset) + J \sin(\emptyset)$$

By assigning values to $\emptyset$, from $\emptyset = 0$ to $\emptyset = 90$ degrees, we obtain both a horizontal and a vertical line, respectively, i.e., the complex plane, which can be "seen" as a fifth sphere of reality or a totality that can be used to represent or contain the four dimensional space-time continuum.

It can also be seen as a mathematical representation of the domain of Form, in the same line of that domain imagined by the perennial philosophy with Plato. Karl R. Popper in the intent to transcend dualism envisioned this domain as a "third world", independent from mind or the subject and the object. Karl R. Popper main concern in <u>Objective Knowledge</u> [10]was precisely to avoid what he called an essentialist explanation by introducing this "third world", as a world independent both from the object and the subject. Only through a mathematical representation we can avoid any semantic pitfall. In the following figure we can see two systems S and S´ in interrelation, and apart each other an angle, being the whole domain of representation the complex plane.

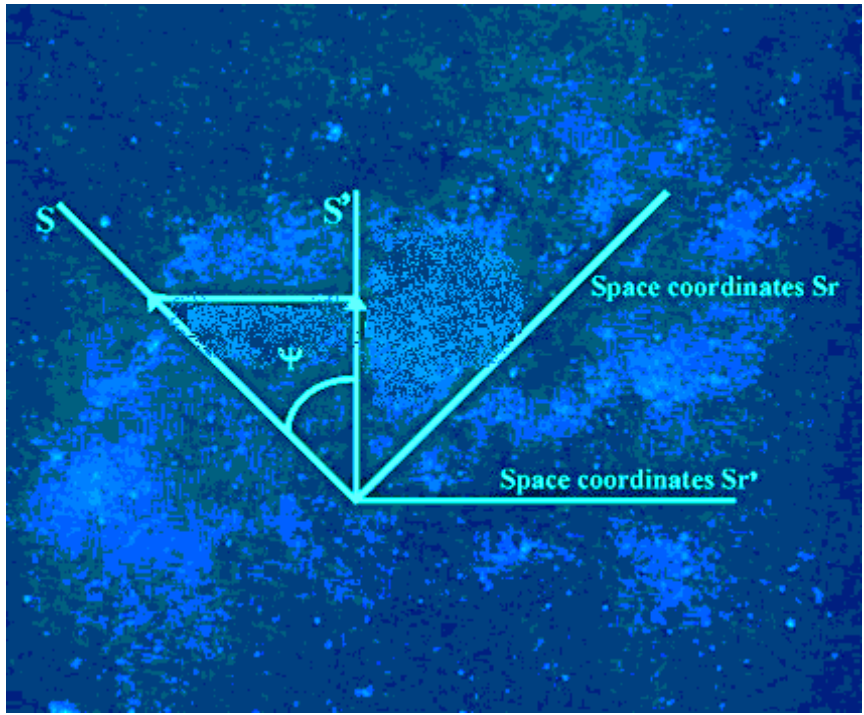

In general we can then envision Ø acquiring all values, from zero to 360 degrees, or just, some sort of a clock pointer, or a vector rotating about an origin at a given frequency, where Ø = wt and w = 2¶f

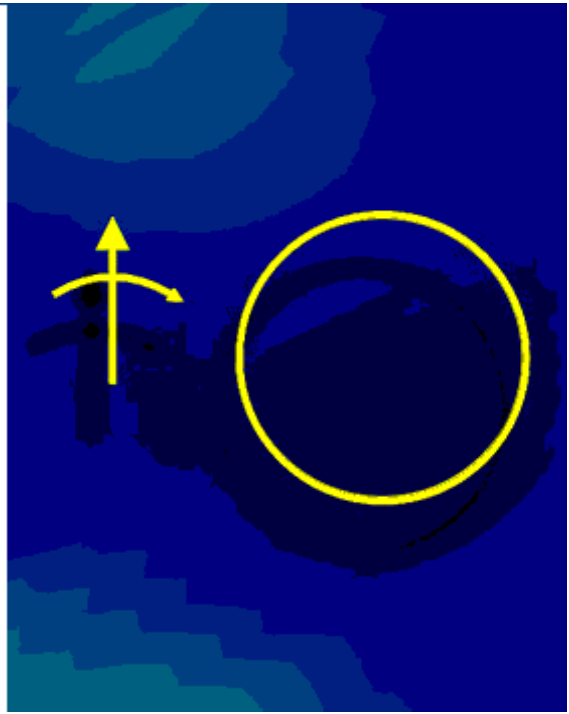

A vector rotating about a point, is what we term the *centerness* attribute of Euler Relation, that has to do with its natural cyclical behavior. We could say nature loves the cyclical waveform behavior, as we find it everywhere, from the motions of the stars and the planets, to the tides on earth, our heartbeats and our psychological states and even that motion that has always exerted such a fascination upon human mind, I mean, the pendulum motion. In Euler relation we have two types of cyclical waves separated by the radical J, and what this means is that those two type of cyclical waves are very different in nature, and this is what we are going to show in the following.

In Euler Relation we then have a:

Dual, symmetric or binary component that can be represented with the sine function:

$$(-) = \text{Sin}(-\theta) \qquad\qquad (+) = \text{Sin}(\theta)$$
$$\underline{\qquad\qquad\qquad\qquad\qquad\qquad o \qquad\qquad\qquad\qquad\qquad\qquad}$$

and a Nondual one component that can be represented with the cosine function:

$\text{Cos}(\emptyset) = \text{Cos}(-\emptyset)$

We must recall that the dual component or that component associated with the sine function changes with changing the sign of the angle Ø, and that the nondual component or cosine function remains the same with changing that sign angle, so we have in this unit those two requisites we pointed out at the beginning of this paper necessary for representing reality. This mathematical process of changing sign is associated with changing the rotation sense, counterclockwise or clockwise as we will see later, so it has to do with a very general sense of rotation of the whole structure.

Historically Euler Relation was associated with the problem of the infinite series:

$$e^x = 1 + x + x^2/2! + x^3/3! + ...$$

where by replacing

$$x = J*Ø$$

we obtain finally Euler Relation by separating those terms that are affected by J from the others, obtaining the nondual or cosine expression and the dual affected by J, or sine expression.

This infinite series is Euler Relation, where, e = 2.71828. The main point to notice at this very moment, is the cosine and sine nature of the two components separated by **J.**

*Cosine or nondual nature of Euler Relation*

From Euler relation we obtain the cosine function represented by

$$Cos(wt) = (e^{J(wt)})/2 + (e^{-J(wt)})/2$$

The cosine function is expressed as the sum of two vectors rotating in opposite directions[14], one of them in counterclockwise or positive direction at an angular velocity w, and the second one in the clockwise or negative direction at an equal angular velocity w. As the vector rotate the two dual components cancel each other, and as so the sum is a purely real or nondual vector, nondual, because we cannot obtain the opposite by changing the sign of the angle. The axis of the cosine function is in fact that axis not affected by J, the so called real axis, we have named the nondual axis instead. It's important to recall the radical duality, expressed in that inherent polarity or tension of those two vectors rotating in opposite directions but in a comprehensive whole, i.e., the complex plane.

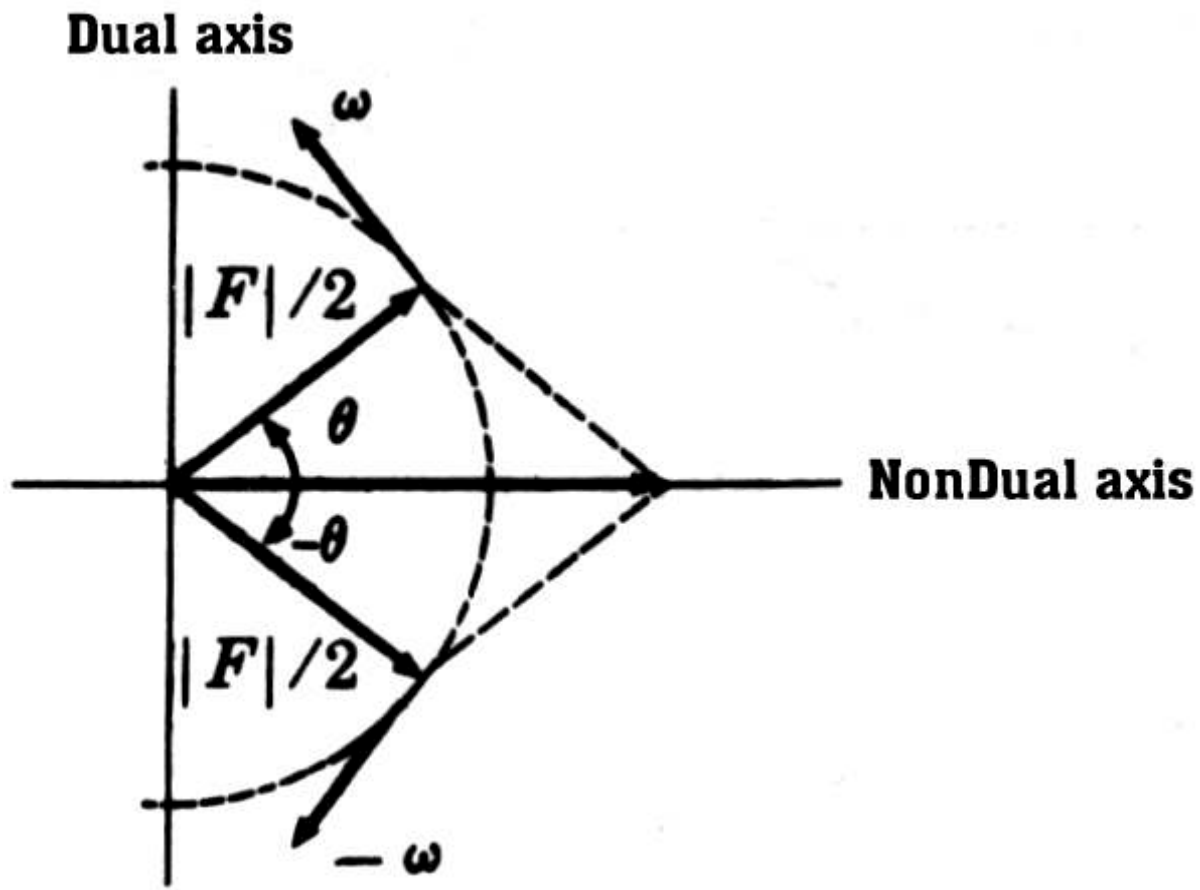

*Sine or dual nature of Euler Relation*

From that same Euler Relation we obtain the sine function represented by

$$\text{Sin}(wt) = [(e^{-J(wt)})/2 - (e^{J(wt)})/2] * J$$

The sine function is expressed as the sum of two vectors rotating in opposite directions, one of them in counterclockwise or positive direction at an angular velocity w, and the second one in the clockwise or negative direction at an equal angular velocity w. As the vector rotate the two nondual components cancel each other, and as so the sum is a purely, as it were, "imaginary" or a dual vector. Dual because by just changing the sign of the angle we can obtain the opposite.

The axis of the sine function is in fact the J axis(see how the sine function is affected by J**)**, the so-called-imaginary-axis, or the symmetry axis, where symmetry is defined as similarity of forms or arrangement on either side of a dividing line, so on one side we have a positive magnitude and on the other we have a negative sign for that magnitude.

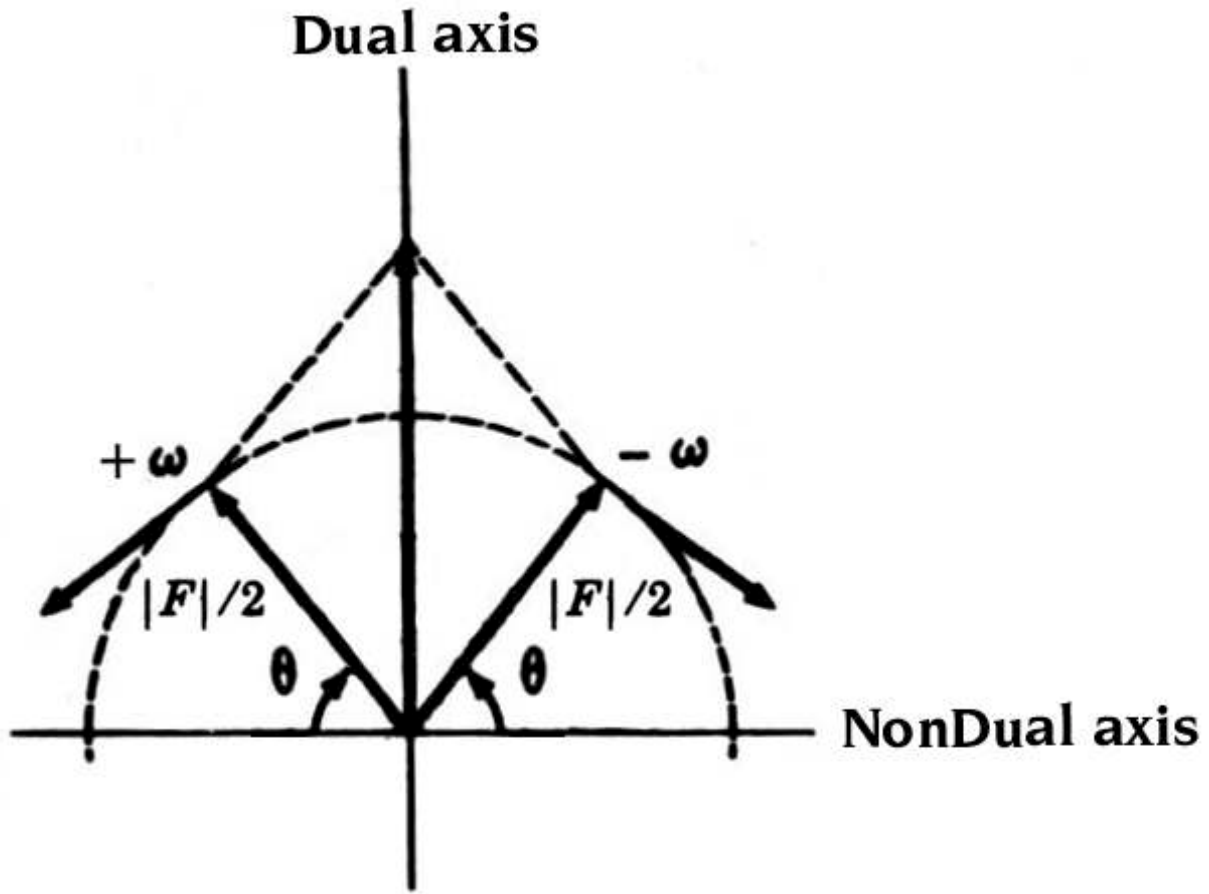

From this point of view it is not perfectly general to say that the sinusoidal wave, as an adjective, is the same for both sine and cosine functions even though we can obtain the one from the other by the addition of a phase angle of -90°; that addition is not at all a trivial one though.

To avoid any ambiguity we must differentiate clearly the asymmetrical or nondual nature of the one, and the symmetrical or dual nature of the other. Reality is in fact composed of two components, but in the physical domain, i.e., the " imaginary" domain, the dual nature is easily grasped and as so we can really say nature "loves" the sinusoid, and normally it hides as a mystery, its nondual nature.

The nondual nature must always be discovered. It is some sort of fixed or nonchanging component to which the change of the system as a whole can be referred, but as a point of reference it must be discovered, it must be chosen or requires a decision. In this sense we can say that the foundation of all reality is an ultimate frame of reference in which the nondual, the nonchanging is at the background[17] or up from the point of view of hierarchy and this implies then an "inclusive" attribute that is essential to have always in mind when dealing with reality. That frame of reference, we have named the domain of Form, permits us to define a domain independent of the observer and the object, as within it the object

**11**

is embedded. We will then be able to have a Quantum Mechanics solution without the "observer drawback", as Karl R. Popper tried to find all his life but from the philosophical point of view and which was Einstein main concern about QM too.

## 5. The resonant effect: "merry-go-round" effect

Phase angle and magnitude are the two main state variables of that unit obtained with Euler relation and that has been named a phasor in EE. The Basic Unit System concept *will be a rotating entity in the complex plane at a given frequency*, that can be codified as:

(Energy, (angle/magnitude))

In most practical problems all vectors, in the complex plane, must be riding with the same frequency, so that a sort of "merry-go-round" effect[14] exists. Vectors are seen as stationary with respect to each other then, so the ordinary rules of vector geometry can be used to manipulate them as with that "merry-go-round" effect we obtain some sort of static framework within dynamism. This is a real impossibility with the monadic "real" numbers mathematical representation.

We have then the chance to have the phenomenon of resonance. We are acquainted with such a phenomenon specially in the production of musical sounds and certain type of vibrations but the important point to recall is that *through resonance we can explain those cases where small changes can often produce large effects*, [13]

We must recall the fact that the phase or the angle sign is associated with the positive or -negative sense of the vector rotation:
- In the cosine case we can interchange both rotation vectors, by changing their sign and nothing changes, the form remains the same, they cannot be segregated just as in a magnet where we cannot differentiate the two polar components as they are seen always as one, as it were, the oneness property.
- In the sine case that interchange, or changing of signs of the angle, affects the final-resulting vector with respect to its position in relation with the nondual axis.

Resuming we have with ER a fundamental structure with two components separated by the radical J**:**
- one nondual in which wholeness is an essential attribute and which brings us to mind that self-awareness capacity that permits us as humans to think about our own thinking process, that permits us reflectivity or to know that we know, a new way of "seeing"; a way to be conscious of our being that makes us different from the animal world. This nondual component is precisely that one that permits us to represent the within of things[4]. With this within attribute we can envision a storing capacity of energy as that one we find in magnetic fields or an information storing capacity in general. *This within attribute as a storing information capacity at its highest manifestation is very different from consciousness***,** being more related with a fifth sphere of reality or just in plain mathematical words with the complex plane. That pretension to reduce everything to consciousness is not our pretension anymore. A greater within means a greater complexity or a greater without, that can or cannot be necessarily a greater consciousness.
- one dual in which we can have two parts separated or clearly differentiated. Parts separated, is precisely that condition necessary for the application of analytical procedures, being the second one, the chance to linearize.

We will see in what follows that ER most important isomorphic property is precisely *to reduce complexity to a minus one degree* making linear the representation of non-linear operators without reducing them. These two basic components are united in that mathematical symbolism called Euler Relation and at the same time separated in a radical way by J, repeating again at a higher level of representation the same basic structure we have found within that same relation, an inclusiveness attribute. We have then a fundamental and basic minimum structural complexity represented by Euler Relation.

Is this basic minimum structural complexity the one necessary to represent reality at its most profound structure? Does this basic minimum structural complexity give us those isomorphic properties needed for representing and deducing the most fundamental laws of physics and reality? The answer to these two questions will be yes, and this is what this proposal is all about.

## 6. Complex Algebra and Isomorphic Properties

Vectors are ideal mathematical entities when relationships are important and as so we have vector sums and differences, but also multiplication, division, derivative and integrals and every one of these operations can be represented in the complex plane as some sort of complex metrics and even the resulting geometrical figures are simpler than those obtained in normal geometry[14]. The powerful advantage of complex number is seen in computations, or when using complex-algebra, where the isomorphic property of Euler Relation manifests all its co-variant power.

The main restriction we must pose from the very beginning is that of the same frequency. All rotating vectors considered must have the same angular velocity or frequency, so that we can have the "merry-go-round" effect[14]. The frequency in this sense is that variable that can in fact produce large effects in a system that was previously chaotic before acquiring it[13]. The order of this system depends on that acquired same frequency for all entities conforming the system. This "restriction" is in fact the central point for the system acquiring *a higher ordered state [13]***.** That frequency is associated with one of the essential state variables of the system, the angle, being the other the magnitude. The state is in fact that one in which having a "merry-go-round" effect gives us as a result an organized complexity.

Rewriting ER again

$$e^{J(\emptyset)} = \cos(\emptyset) + J\sin(\emptyset)$$

we can note that it is a unit vector standing at an angle Ø from the nondual or real or main axis

$$1 = \text{Sqr}[\cos^2(\emptyset) + \sin^2(\emptyset)]$$

that can be represented by the following figure

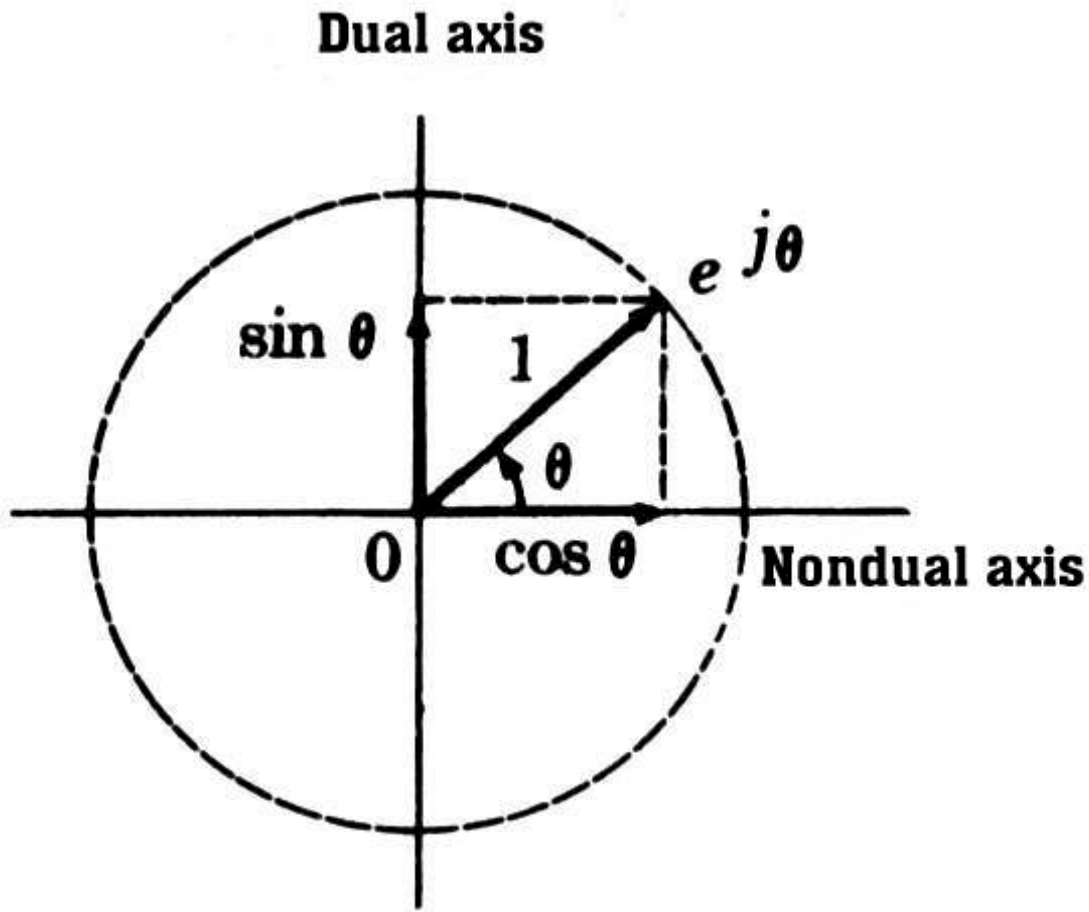

We can have then in general, entities represented as

$$A = \text{Abs}(A) * e^{J(\emptyset)}$$

that can be represented in a rectangular form as

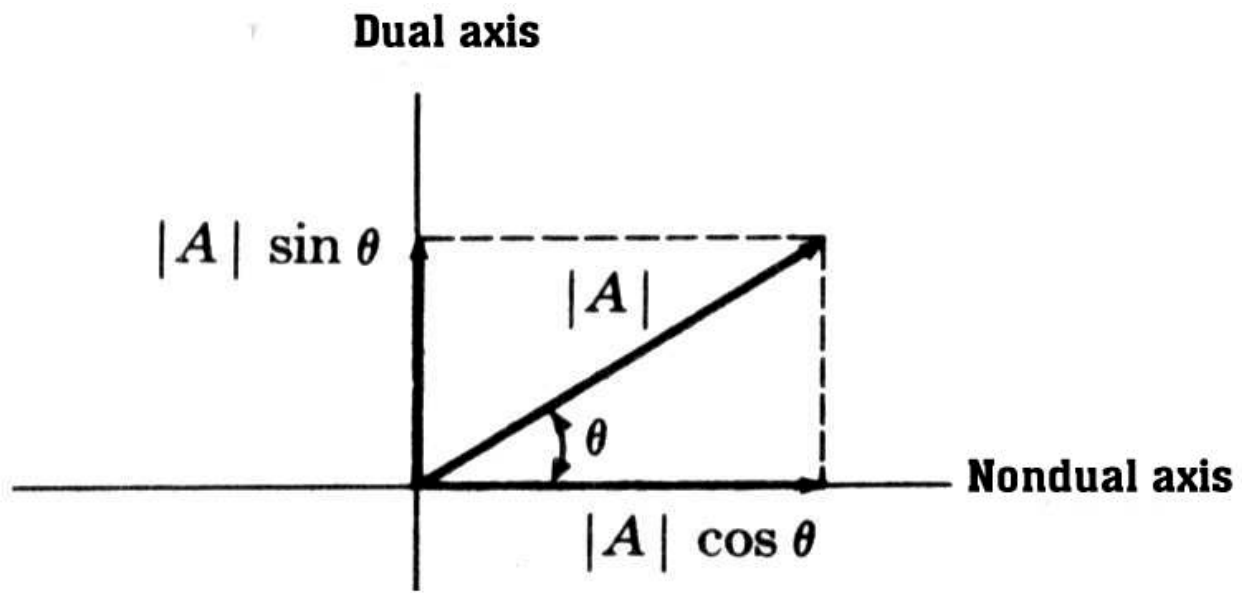

or

A = a-nondual + J * a-dual

where

A-nondual = Abs(A)*Cos(Ø)

And

A-dual = Abs(A)*Sin(Ø)

With these two expressions the complex algebra was established having in mind that we must not mix oranges and apples. Must we abandon the reduction tendency with this complex representation?. Must we make a radical distinction between the two basic components that are used to represent reality?.

The nondual components must be used with nonduals, and the dual components with duals. In this way all laws of normal algebra and arithmetic are preserved. This radical separation preserves homogeneity on the one hand and heterogeneity on the other.

If we call qualitative those aspects related with the phase angle it is shown they are summed instead of multiplied within complex algebra. This capacity of complex numbers to reduce a nonlinear operation to a linear one, is in fact the one that makes them such a powerful simplifier tool, and adequate to represent the complex nature of reality. Linearization does not affect at all the quantitative part, just the qualitative one. This gives us *a new capacity to think the whole correctly* and the chance to introduce boldly in our intellectual frameworks new categories. A new world outlook is obtained that includes in itself an explanation and development of new things and even solves once and for all the paradoxes of wave and particle as we will see when deducing the complex Schrödinger Wave Equation.

*Differentiation and integration.* We have already pointed out that remarkable property of ER we have qualified as isomorphic and that has to with the fact that the integration and differentiation of ER, are themselves ER of the same frequency.

By the rules of elementary Calculus, the derivative of an ER with a constant magnitude has the same magnitude as the original vector multiplied by the frequency w, but the angle has been advanced by 90°.

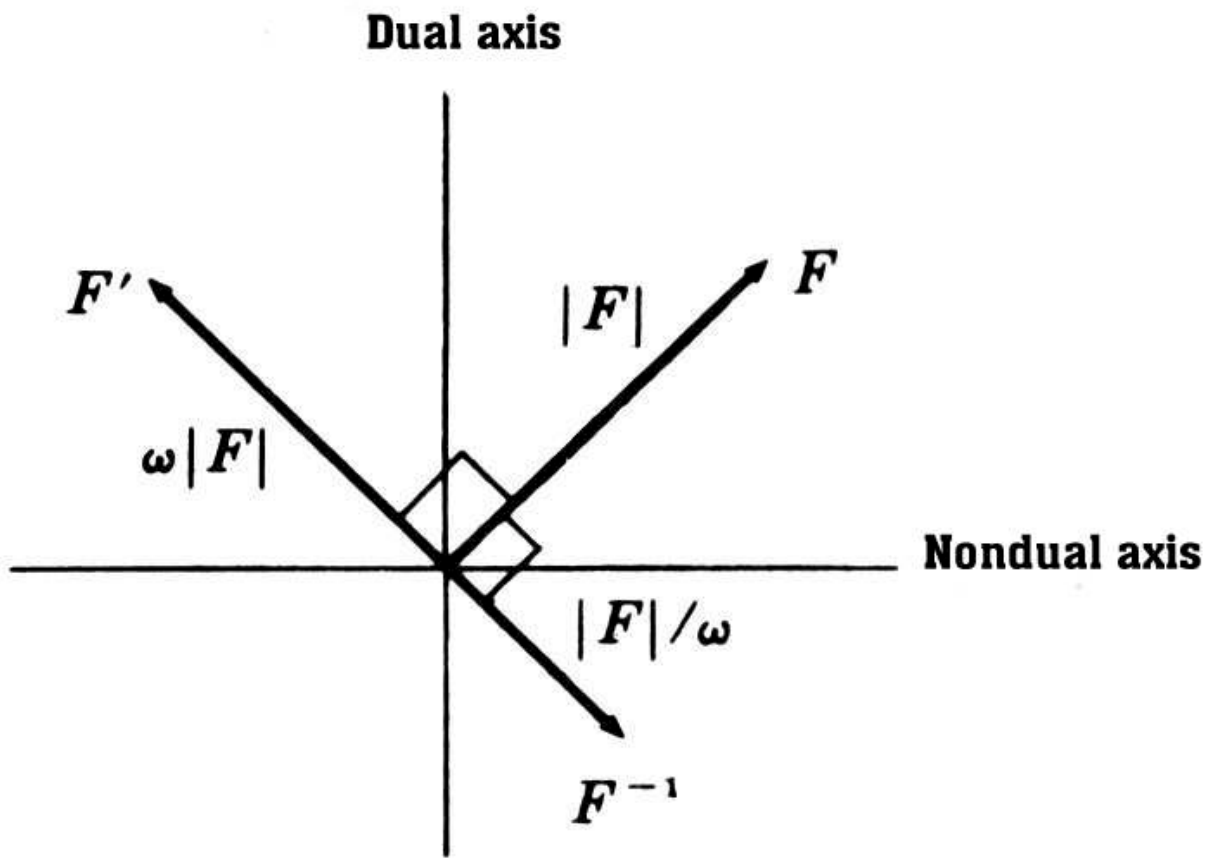

And the integration is the dual operation as that of differentiation and as so the integration of an ER with constant magnitude, has the same magnitude as the original vector divided by the frequency w, the angle has been retarded by 90°, or just the expression multiplied by -J , as were, an opposite sense of rotation.

$F^{-1}$ (INTEGRATION)

$F'$ (DIFFERENTIATION)

This way of reasoning is exactly the application of what can be named the duality principle, which is a way of applying a binary logic or symmetry to obtain a dual reality from the other components.

The complex vector algebra simplifies mathematical operations by one degree or it reduces complexity by one degree. With it a real isomorphic tool is obtained as even complex operations such as differentiation and integration are reduced to simple algebraic operators.

This reduction of complexity by minus one degree permits us to comply with those two conditions of analytical procedures:
- linearization and
- to have weak interaction between the parts in those cases in which the general open system is closed, which was what Ludwig Von Bertalanffy pointed out as the central and methodological problem of systems theory[12].

Euler Relation is the ideal isomorphic unit, as the sums, differences, integrals, and derivatives of Euler Relation functions of a given frequency are themselves Euler Relation functions of the same frequency, they do not change their form with these operations, they remain invariant or just co-variant. No other mathematical function is preserved in this fashion, as so it is the ideal one for filling that requirement Einstein put in his <u>The foundation of the General Theory of Relativity</u>, [5], when he wrote "The general laws of nature are to be expressed by equations which hold good for all systems of co-ordinates, that is, are co-variant with respect to any substitutions whatever(generally co-variant)".

**7. A complex Geometry and Synergy**

Our next step is to define a differential complex geometry in which we have a differential of reality defined as a Basic Unit System or a Holon such that

$$DS = Abs(DS) * e^{J(\emptyset)}$$

The late Abraham Maslow was the one who coined the term "synergy", an obscure term from anthropology, but he used it for the first time in business to describe how wealth can be created from cooperation. Creation of wealth, emergence of new things and structures that modify themselves to give better performance are main issues of the so-called-sciences-of-complexity or systems sciences[13].

Structure and environment or thirdness(Oneness)[16] and openness they both claim for a whole that is greater than the sum of the parts, for a basic and fundamental framework in which three fundamental and interdependent entities are put in mutual interaction. They pose the need to have as starting point not only a minimum structure, but also a minimum system of elements in interaction.

In the physical domain synergy can be found whenever we have a real transformation of energy to another useful way or level which presupposes interchange or transformation of energy between systems. This "useful way" is in fact the real source of new applications and in this sense synergy is there, where we have emergence of new realities or just open systems that have exchanging-energy-capacity with the environment.

That whole greater than the sum of its parts has nothing to do with metaphysic, as that "greater" comes from the very nature of an open system.

Electrical energy, or the alternating current we use at home, is in fact the result of the application of the Principle of Synergy where energy is transformed from a primary source, hydraulic to electrical energy, by moving a threefold magnetic structure, which gives at the output the AC energy we can utilize in many ways.

If we consider three-double-complex-vectors or just Buses, one in space and one in time, codified as:

$$A = Abs(A) * e^{J(\theta)} + e^{J(wt)}$$

$$A = Abs(A) * e^{J(\theta+120)} + e^{J(wt)}$$

$$A = Abs(A) * e^{J(\theta+240)} + e^{J(wt)}$$

where they correspond to a three physical disposition displaced 120° from one another, on the one hand, and on the other to a dynamic entity or field. It is shown that the physical disposition cancels or the cause-effect relationship is subtle. It depends on the interchanging of energy or information with the environment through field concept.

At last we have

$$A' = Abs(A') * e^{J(wt)}$$

a rotating entity or Bus at a given frequency.

*A given frequency* is precisely the one attribute that makes it possible, the "merry-go-around" effect, i.e., that new order or reality that comes from the application of the Principle of Synergy. If the frequencies are not the same for the three independent entities we won't have the desired "merry-go-around" effect. We have some sort of emergence state in which one small effect -a small change in frequency- gives us a large effect.

Those three, or six independent entities, in interrelationship conform one entity, a whole picture, a form represented by a circle.

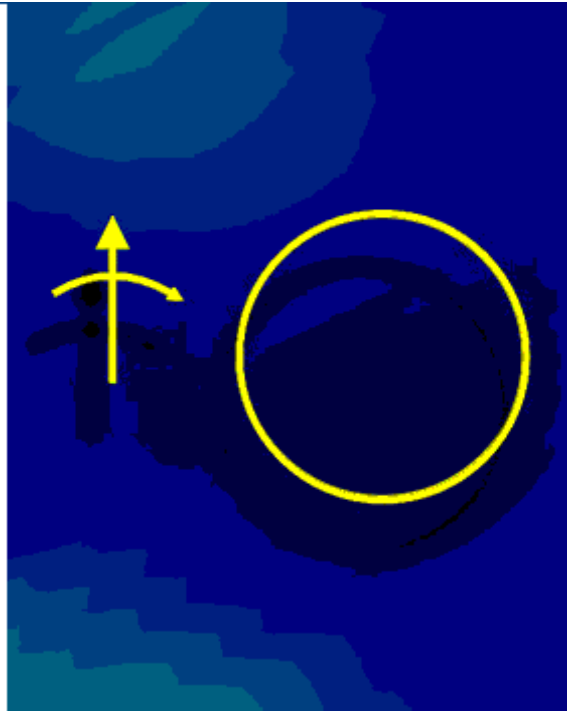

The Principle of Synergy, or the gluing principle can be represented by a fundamental sevenfold structure

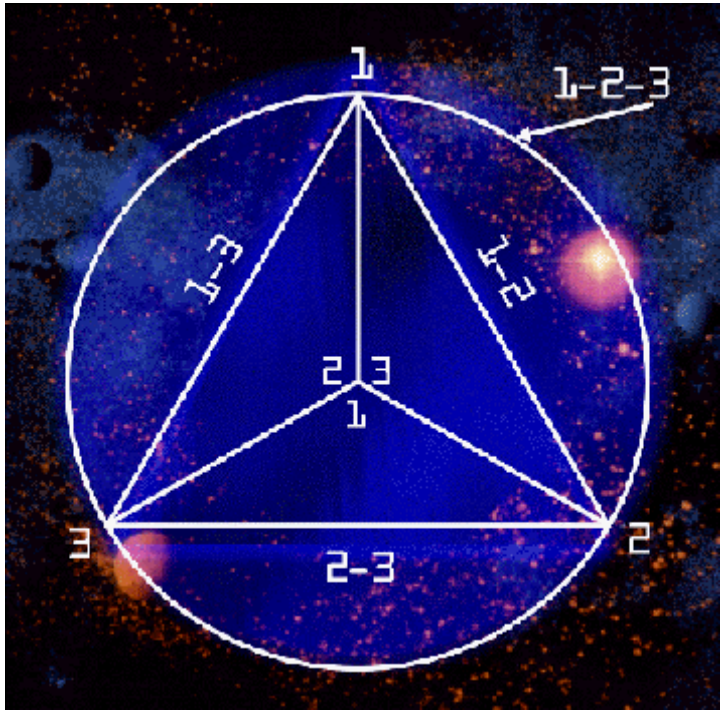

The main point to recall is the structure represented by
- three inner or nondual relations, or just the 1-1, 2-2 and 3-3 relations
- three outer or dual relations, or just the 1-2, 2-3 and 1-3 relations

The wholeness attribute is associated with an emergent property represented by the 1-2-3 relation.

Even though we have a holistic property in this codification, it is embedded in a Basic Unit System or Holon, not in a general abstraction where the hierarchy-relationship between the part and the whole is not clear at all.

The Bus(Holon) concept is embedded in a formal complex mathematical framework, whose generality is associated with those isomorphic properties we have already presented and that gives us a real internal consistency which allows to treat steady-state systems by the same general techniques or methodology.

Furthermore those six relations are some sort of detailed complexity as opposed to the 1-2-3 relation that is the complexity of the whole. The traditional problem between the part or detail and the whole is not a problem anymore with this sevenfold structure that can be used at different levels of reality without reducing the one to the other, as its "actualization"[16] depends on a frequency, not necessarily in a law or relation between the two basic components.

The fact we use a complex mathematical symbolism, as a tool, for representing this sevenfold structure avoids that syncretic whole reasoning Galileo used when he wrote *There are seven windows given to animals in the domicile of the head, through which the air is admitted to the tabernacle of the body, to enlighten, to warm and to nourish it. What are these parts of the microcosmos?*

*Two nostrils, two eyes, two ears, and a mouth. So in the heavens, as in macrocosmos, there are two favorable starts, two unpropitious, two luminaries, and Mercury undecided and indifferent. From this and many other similarities in nature, such as seven metals, etc., which is were tedious to enumerate, we gather that the numbers of planets is necessarily seven.* But it also prevents using as a key guiding principle *the primacy of the whole* which results in arrogant role of dominance when applied to organizations. But let's go to the applications.

## 8. The Pendulum

The pendulum movement is so remarkable not just for its role as a cornerstone in the birth of modern science that permitted Galileo a paradigm shift, but also because, it is the most natural example of an open system. When looking at the swinging body he saw a body that almost succeeded in repeating the same motion over and over ad infinitum. Its success in repeating the same motion over an over lies in its interrelationship with the environment or on its openness attribute.

By looking at the pendulum *Galileo reported that the pendulum's period was independent of amplitude for amplitudes as great as 90°, his view of the pendulum led him to see far more regularity than we can now discover there.[18]* Is this view of seeing more regularities than there really existed part of the incapacity of normal science to explain pendulum-like regularities in a most documented way, and following a mathematical methodology with no leaps? Or *How else are we to account for Galileo's discovery that the bob's period is entirely independent of amplitude, a discovery that the normal science stemming from Galileo had to eradicate and that we are quite unable to document today. [18]*

The exact simple pendulum solution implies the solution of a first order differential equation which implies too an integration whose solution is an elliptic integral. This means the introduction of an approximation factor that could only be found by observations of the pendulum real behavior, some sort of trial and error procedure. This normal mathematical symbolism, I mean, not-complex-mathematical symbolism, cannot give reason of that approximation factor without using some sort of methodological leaps, to explain deviations.

Normal science works with closed systems, that is, systems that do not exchange with their environment. The pendulum movement seems to violate even that infamous second law of thermodynamics. It gives us a natural sense of eternity just as the poet *Jacques Bridaine* wrote

*"Eternity is a pendulum whose balance wheel says unceasingly only the two words, in the silence of a tomb, 'always! never! always!...'"*

After exploiting the cyclical wave nature of Euler Relation in EE, it is obvious to expect we will be able to explain all those natural phenomena such as that of the pendulum, in which we have cyclical or wave movements too.

For achieving this it is necessary to realize a real paradigm shift. With the complex plane we have introduced in fact a new sphere of reality in which we have embedded both the nondual and dual nature of reality, and as so a complex metrics, whose main characteristics are:

- on the one hand, its isomorphic property that gives us a powerful methodological tool to explain those cases in which real dynamism is involved
- on the other that property we have associated with a "merry-go-round" effect that permits us to make a mathematical representation of real dynamic entities or else generation of forms. In the pendulum its form is continually generating itself by just a mere impulse.

A new sphere of reality or the sphere of form, in which the four-dimensional space-time continuum is embedded. A new category or totality that contains the physiosphere, but also that sphere in which we can have animated forms, as it were, the biosphere, or the Bergson-Theilhard de Chardin DURATION[1,4] concept which is quite different from the space time continuum.

A complex metrics is then a top-down metrics. From the notion of the Basic Unit System and those normal regularities already known from physics, but also from those properties obtained from geometry, we can obtain the state of our Bus system, through a methodology in which instead of starting with the postulation of a differential equation, we start with the Bus concept as a tool to integrate, or obtain that state.

From the point of view of the BUS, the pendulum movement is a rotational motion. The pendulum as a Bus is then an open steady state system. The earth gravitational field is its context.

By observing that cyclical movement we can observe a maximum angle θ, or θmax, for a corresponding maximum displacement Smax.

Solving this problem means, integrating , some sort of constitutive characteristic, not a summative one.

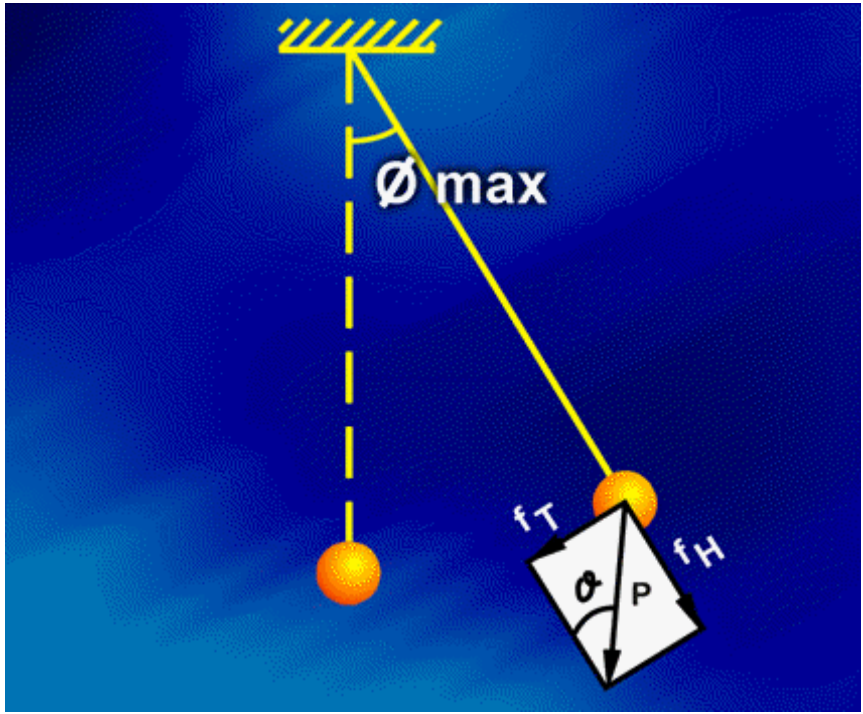

The complex trajectory DS between S = o to S = Smax will be

$$\int_0^{Smax} DS = \cos(\theta max) * \int_0^{Smax} DS + J*\sin(\theta max) * \int_0^{Smax} DS$$

so we have

$$S = Smax * e^{J(\theta max)}$$

we note that this is a static and potential expression related with space, and as so we must introduce its dynamic counterpart by multiplying both parts of the equation by the basic unit of time given by:

$$e^{J(wt)}$$

where θ = wt

The reality S of the pendulum is generated and represented by:

$$S(t) = S_{max} * e^{J(wt + \theta max)}$$

Intuitively, from geometry we know that in a circle we have

$$S_{max} = L * \theta max$$

Where L is the radius or the cord length of our bob system and θmax, that angle subtended by the arch, so we have

$$\mathbf{S(t) = L * e^{J(wt + \theta max)}}$$

as a general dynamic expression in which the principle of synergy is included. Whenever we have rotation synergy is at hand.

## 9. Balancing Equality and the Pendulum

To obtain the pendulum harmonic motion or just its steady state, we must apply a balancing or a compensating equality, between two driving forces or polarities:
- an inner one related with the weighting mass of the bob and
- an outer one related with the so-called-inertial mass.

This equality played a very important role in Einstein's works about gravitational fields[5]. Up to that time that equality had been considered a mystery in the Newtonian framework. It generated those famous thought experiments with elevators, in which bodies are subjected on the one hand to an inertial force, and on the other to the gravitational field.

When these two forces or polarities within a comprehensive whole become equal then we have a harmonic motion or a motion that almost succeeded in repeating itself over and over ad infinitum. A balancing loop which implies a rotational motion or the Synergistic Principle means that the law of opposites must always be considered in a comprehensive whole, or dynamic context, producing a cyclical movement. But if we consider inertial forces then we must apply Newton Second Law, but this time in the fivefold continuum, or in the complex plane context.

The first derivative of S(t)

$$S(t) = L * e^{J(wt + \emptyset max)}$$

is

$$dS/dt = L * \emptyset max * e^{J(wt + \emptyset max)}$$

and the second derivative:

$$d^2 S/dt^2 = -(L*Ømax*w*w)*e^{J(wt + Ømax)}$$

being

amax = ( L * Ømax * w *w *)

the maximum absolute value of acceleration.

According to that second law we have at the maximum point Smax the inertia "vector", which tends to maintain the motion as:

**Fmax = m *( L * Ømax * w * w)**

if we apply, on the other hand, the vector composition of forces and using tangential components at that point we have the mass "vector" as:

**Fmax = m * g * Sen(Ømax)**

so equating both the inertial and the mass "vectors" we've got:

L*Ømax*w*w = g*Sen(Ømax)

the angular velocity is:

w = sqr( g * Sen(Ømax) /( Ømax * L))

but if

w = 2*Pi *f

and

T = 1/f, where f is the frequency and T the period then

T = 2*Pi *sqr((L/g) *( Ømax /Sen(Ømax))

which is the exact pendulum formula.

The factor

sqr( Ømax /Sen( Ømax) )

tends to one when Ømax is small and it can be omitted as long as the amplitude does not exceed 10°.

In Table 1, in the second row we can see the values K of the elliptic integral, in the third row the corresponding factor according that integral, and in the fourth row the corresponding factor obtained with the Bus concept. In the fifth row we can see the difference in error between the elliptic factor and the Bus Factor. This difference in errors as the amplitude increases can be taken as a fallibility criterion for those angles greater then 30 degrees where the error between the two is definitely unacceptable.

**Table 1.** Approximation Factor for the Period of a simple Pendulum

| Ø(angle) | 0° | 10° | 20° | 30° | 60° | 90° | 120° | 150° | 180° |
|---|---|---|---|---|---|---|---|---|---|
| K | 1.571 | 1.574 | 1.583 | 1.598 | 1.686 | 1.854 | 2.157 | 2.768 | Infinite |
| 2K/Pi | 1.000 | 1.002 | 1.008 | 1.017 | 1.073 | 1.180 | 1.373 | 1.762 | Infinite |
| Ø /Sen(Ø) | 1.000 | 1.005 | 1.002 | 1.047 | 1.209 | 1.570 | 2.418 | 5.235 | Infinite |
| Error(%) | 0.000 | 0.298 | -0.595 | 2.865 | 11.248 | 24.84 | 43.217 | 66.34 | N/A |

The exact solution introduces a nonlinear character in the pendulum motion which gives reason of its real behavior defined by the well known observed laws. Normally this problem was solved by resolving a differential equation, that gave finally an elliptic integral, or some kind of solution that must be validated with the real observed behavior of the pendulum, as an elliptic integral cannot be expressed in terms of the usual algebraic or trigonometric functions, see <u>Vector Mechanics for Engineers</u>. [3]

Galileo did not know this factor, and so he extended the application of the observed pendulum laws far beyond its real range, even to 90 degrees. But the fact that the time of any amplitude were independent of the mass of the body made he think in the falling of bodies and specially in the well-known Pisa Tower experiment, in which both an iron and a wood sphere fall with the same time, which means in all cases, the trajectory followed by those bodies is an invariant.

The pendulum is essentially a device for measuring time. It is in fact the contrivance of time. Its form unfolds itself in time, it is the dynamic device for excellence. Differently from those movements of classical Newtonian mechanics, like those of the planets, in which time does not play really any role and in which we could talk about the chance to invert time, the pendulum movement in this sense is not a classical movement, and it answers the question posed by Thomas S. Kuhn about that movement.

The Principle of Synergy applied to the pendulum explains why this principle seems opposed to that infamous second law, as with it forms can be generated and we can obtain in natural way a steady state open system.

# 10. Quantum Mechanics

The Quantum Mechanics problem before being a scientific problem is a philosophical problem as from the beginning it touches the same nature of reality and the relation of the subject with that reality. The relation between object and subject, the active factor of the subject in the process of cognition is in fact an epistemological problem that has to do with the way the subject defines what he understands by objects[10, 15]. But this philosophical problem can be found in the idea of the active role of language in the shaping of our worldviews or images of the world, and as so it is primitive and was the main concern not only of philosophers but also of historians as we can see in Adam Shaff's book History and Truth. And we know also the efforts done by Karl Popper during all his life and specially in the introduction of his <u>*Quantum theory and the Schism in physics. From the Postscript to the Logic of Scientific Discovery*</u> [11] to exorcise consciousness or the observer from physics.

Reality is independent of human mind, of the subject perceiving it, and we can distinguish in it three great realms:

- the subject that perceives that reality

- the object perceived by the subject as reality, the intersubjective domain

- reality per se

that we can codify in a holonic way as

(The subject, (the field of consciousness /reality)

or just as the big three

("I", ("We"/"It"))

The structure of reality perceived by the subject is crucial in the determination of the objects of study, those objects normally studied by science. But science is a product of human activity and as so language is crucial in the determination of those objects of study. This problem took Karl R. Popper to define a third world, as an intermediate world between consciousness and objective reality[10].

As we have seen we can have two kinds of mathematical languages, a partial one that not necessarily denies complex numbers as it uses them in "convenient ways", and an integral one in which complex number are used for building a unit in which the radical J is some sort of operator to distinguish between two different orders of reality, as it were, the nondual and the dual nature of reality embedded in that same unit. In this sense this is not a new theory of everything but a way of "seeing" and interpreting reality in which uncertainty is always concomitant.

When we talk about Reality per se we are not meaning a third world as that of Plato, divine, superhuman and eternal, but of reality as defined by Mortimer **J.** Adler in his <u>Adler's Philosophical Dictionary</u> **[9]** where he wrote:

*BEING The word "being" is an understanding of that which in the twentieth century is identified with reality.*

*What does the word "real" mean? The sphere of the real is defined as the sphere of existence that is totally independent of the human mind...Another distinction with which we must deal is that between being and becoming, between the mutable being of all things subject to change, and the immutable being of that which is timeless and unchangeable. That is eternal which is beyond time and change. In the real of change and time, past events exist only as objects remembered, and future events exits only as objects imagined.*

Historically the prevailing structure of reality has been a dualistic one in which reality as a whole is divided in just two domains, being each one, as it were, parallel to the other, with no possibility of integration in case of gross dualism. Normally there has been some sort of partialness depending on the accent put either in philosophy as the queens of all science, or in science as the real-objective science. Philosophy and Ontology are not considered normally as independent domains, but for our purpose and aim, philosophy must look for an integrated image of reality and in the searching of methods of generalization so that philosophical theses are kept from contradicting science, in other words it must contribute to oneness and not to partialness. On the other hand Ontology must look for the more-being, and as such realism must be concomitant as a main issue.

An open system can have a variable center and as such its external manifestation or its form cannot be determined, as its field or circle of influence is variable, and this will be the case for the electron. Other open systems in which both its center and its radius are not variable will have a more determined state and will be near to an ideal closed system. But this issue has to do directly with the Uncertainty Principle, in which we have as in philosophy no univocal conditions as those we can obtained with closed systems. But the important point is the correlation we have between closed systems, measuring instruments and open system and the Uncertainty Principle. The incapacity to measure the momentum and the position of an electron at the same time, has been associated with a philosophical problem that has to do with an emphasis on the subject doing the experiment, pointing out the fact that the structures of the subject are crucial in the way he defines reality or its object of study, which is similar to that philosophical problem in which language is considered as a factor that creates reality.

We can talk about metaphysics when we have assertions for which we don't have procedures on the one hand, and the laws on the other, that can be used to verify those assertions. In the latter case when we have laws we can built measurement instruments to verify the Data. On the other hand we have known procedures that when used they take us to a series of consistent results, that can be described qualitatively, but not necessarily quantitatively. In both cases our main task as scientists and philosophers must be the searching of truth.

In our case we will consider this problem of the Uncertainty Principle, in case of an electron, as the impossibility to determine the state of the corresponding Basic Unit System because there is not a law with which we can interrelate its two components, as it were, its dynamic counterpart and its static one. In this way we can separate the scientific problem of Quantum Mechanics from a philosophical problem which I think was the aim of Karl Popper when defining his "third world theory". In our case there are clear differentiation between precision and uncertainty, as there are cases such as the electron case in which there is no way to reduce the qualitative to the quantitative, which was Karl R. Popper requisite to increase the degree of contrastability of certain theories.

## 11. The complex Schrödinger's wave Equation

The trajectory or more appropriately the reality of an electromagnetic entity as that of the electron, which as a matter of fact behaves as a wave, can be represented by the bus concept as

$$DS = Abs(DS) * e^{J(\emptyset)}$$

we can codify it too as

(energy, (time/space))  or  (energy, (t/s))

or

(Energy, (wave/particle))

A general conceptualization of the Bus concept can be seen as a mathematical codification of energy in its most primary definition, but energy "is like a frequency multiplied by Plank's constant h"

$$E = h*f$$

and that angle ø can be replaced by

ø = 2*p / h * (p* x - E *t)

a well-known expression taken from Quantum Mechanics

and

1/λ = p/h

where λ is de Broglie's wave length, h is Plank' s constant, and p is the momentum of the particle –a BUS rotating at a very slow unknown rotational speed in the complex plane- that can be written too as

$p^2 = 2*E* m$

The two state variables of the BUS in this case are x and t, so to able to determine its state, we must find or know a relationship between them. The geometric-like behavior of the pendulum gave us the clue to find that relationship. In case of the planet movements it is the second Kepler's Law associated with a central force movement the one that permits to determine its state and in case of the Lorentz's transformation group, the fundamental equation of electromagnetism it was the constancy of the velocity of light[6].

The electron does not behaves like a classical particle or a planet subjected to a central force, we don't know, like in the pendulum a way to link its two components, as it were, its energy and its particle nature. Its state cannot be determined.

For the electron we cannot find a relationship between its two state variables, being this another way to present the Uncertainty Principle which means from the geometrical point of view we will not have a known trajectory followed by the electron, as was the case for the pendulum and the planets. In both of these cases we have a "real" differential equation associated with. In the electron case we have the well-known complex Schrödinger wave Equation that was presented by him in 1926 as a postulate, i.e., as a way of saying that equation does not have just one solution starting from initial conditions, so the need to measure was replaced by qualitative methods, where one must focus in a behavioral area and not in finding laws that permit us to measure in the classical way. In case of the pendulum its form changes with time, but at the same time, its centerness attribute is localized, which seems not to be the case for the electron case.

Our aim is to deduce that complex wave Equation in the context of the Bus concept. Let us suppose an unknown general solution, as a function of space and time, as it were, of its two state variables

$S(x, t) = Abs(S) * e^{J(2\lambda/h*(p *x - E *t))}$

Please note the variable Ψ(x, t) usually called "the wave function" in QM has been replaced by S(x, t) which is a complex function that contains by definition a wave character[7]. S(x, t) has to do with the real trajectory of the entity in question in the complex plane so in a certain real sense it is a complex amplitude too.

Let us rename

## Abs(S) by S

to distinguish the magnitude from the complex quantity. Expressing E as function of momentum and replacing

$E = p^2/2m$

we have

$$S(x, t) = S * e^{J(2p/h*(p*x - p^2/2m*t))}$$

$$S(x, t) = S * e^{J(2p/h*(p\,x))} * e^{-J2p/h* p^2/2m *t} \quad (1)$$

the point here is to follow the well-known-wave procedure by making two partial derivatives of this expression with respect to space and one partial derivative with respect to time. By equating them we obtain the complex Schrödinger Wave Equation. At this point it is important to recall the isomorphic property of Euler Relation as the one that makes it possible this result, as it were, the permanence of the unstable.

The first partial derivative with respect to space gives

$$\partial S(x, t)/ \partial x = J(2\pi/h)*p*S*e^{2\pi J* p /h* x} * e^{-2\pi J/h*( p^2/2m)*t}$$

where the internal derivative of

$e^{2\pi J* p /h* x}$

is $J(2\pi/h)*p$

and the rest of the expression can be represented by $S''$ so

$$S'' = S * e^{J(2\pi/h)*p} * e^{-J(2\pi/h)*(p^2/2m)*t}$$

the second partial derivative gives us

$$\partial^2 S(x, t)/\partial^2 x = (J(2\pi/h)*p)^2 * S''$$

or else $S''$

$$S'' = \partial^2 S(x, t)/\partial^2 x / (J(2\pi/h)*p)^2$$

$$S'' = -\partial^2 S(x, t)/\partial^2 x * h^2 / 4\pi^2 * p^2 \qquad (2)$$

If we now take the partial derivative of (1) with respect to time, the internal derivative is

$$-2\pi J/h*(p^2/2m)$$

so we have

$$\partial S(x, t)/\partial t = -2\pi J/h*(p^2/2m)*S''$$

or else

$$S'' = \partial S(x, t)/\partial t / -2\pi J/h*(p^2/2m)$$

$$S'' = -\partial S(x, t)/\partial t * 2hm / 2\pi J * p^2 \qquad (3)$$

equating both Ψ'' expressions 2 and 3

$$\partial^2 S(x, t)/\partial^2 x * h^2/4\pi^2 * p^2 = \partial S(x, t)/\partial t * 2hm / 2\pi J * p^2$$

and by eliminating and organizing terms on both sides we obtain finally:

$$\frac{\partial^2 S(x, t)}{\partial^2 x} * \frac{h^2}{8\pi^2 * m} = \frac{h}{2\pi J} * \frac{\partial S(x, t)}{\partial t}$$

we obtain the well-known Schrödinger's Wave Equation, introduced by him in 1926, for a free particle moving in x' s direction. This equation was presented then, as a postulate as there was no means to deduce it, from more basic principles, none the less, we have just applied the wave procedure to the BUS concept based on the principle of synergy. Richard P. Feynman wrote[7] in his famous lectures on Physics:

*In principle, Schrödinger 's equation is capable of explaining all atomic phenomena except those involving magnetism and relativity.*

We must take into account though that the Bus concept as a powerful tool has been used for finding the Lorentz's group of equations that give reason both of relativity and electromagnetism, but also of gravitational fields.

The symbolic representation

(Energy, (t/x ))

establishes a hierarchy in the sense that wholeness or energy can stand on its own, and the other two state variables, t and x, cannot. When the state of the system is determined t and x are related by means of some law or relation, the wholeness or energy is linked to (oneness, openness). When they cannot be linked as the case of the electron we have uncertainty and energy prevails on its own.

There is an emergent conclusion in all this, and it is that ***the fundamental of physical reality is energy***, and not a particle( or the mass concept).

A particle is a Bus or Holon rotating at an unknown low frequency but in the complex plane or in reality represented in that plane, so that its state is completely determined, or else the qualitative aspect is reduced to the quantitative one in such a way no possibility of a field or just a storing capacity is feasible different from its well determined state. It is important to remember at this point that the Bus concept as a mathematical symbolism also has that wholeness attribute that is at the base of growth, each whole has the capacity to generate another whole, but always having in mind that openness attribute that also permits us to define the Bus in general as a symbol for representing a steady state open system. It is not a metaphysical concept that came from nothing. It is a concept in which a gluing principle as that of the Principle of Synergy permits us to define that oneness attribute.

## 12. Conclusions and Suggested applications

From all this exposition emerges a conceptual framework that not necessarily reduces everything to the most elementary levels of reality, but as a good engineering conceptual tool opens new avenues for future research and development. The important point to recall at the outset is the need to abandon the old dualistic framework that has the natural tendency to put the whole or the part as a "primacy" not as a fundamental component of the Basic Unit System. The Bus concept is a part/whole complex mathematical concept that has embedded, as we have seen, the nondual and the dual nature of reality, but also the Principle of Synergy in which the whole is greater than the sum of its parts that gives us a medium to interpret reality, as it were, in organic ways. We have a minimum threshold of complexity in that structure.

In general, reality can be analyzed as buses within buses within buses, but the Principle of Synergy implies a *same frequency* to obtain that "merry-go-around" effect or that emergent steady state or new order or that higher complex state where we have an organized complexity. One of the most important point to recall of this symbolism is that we do not need to make any reference to anthropomorphic concepts such as psyche or consciousness to explain those emergent states where we find a whole greater than the sum of the parts. But from the point of view of the whole reality we obtain a framework in which by the introduction of a fifth sphere of reality, that of form, we have then a sevenfold structure where the space-time continuum is embedded in that fifth sphere, where life can be defined as an animated form, but then we have mind or the noosphere as the sixth sphere of reality, but also the Being as that one that has embedded them all, in an all encompassing way.

Having found such a powerful explaining tool, it is obvious to feel an imperious need to share with the scientific community such a framework that can not only illuminates its actual practice by defining the objects of study in certain cases but also establishing clearly the impossibility to reduce those objects to isolated units in other cases. The whole concern of Ludwig von Bertalanffy in its GST[2] when he wrote

*We may state as characteristic of modern science that this scheme of isolated units acting in one-way causality has proved to be insufficient. Hence the appearance, in all fields of science, of notions like wholeness, holistic, organismic, gestalt, etc.,*

*which all signify that, in the last resort, we must think in terms of systems of elements in mutual interaction.*

is then the same concern of all this paper, but our main point is the integration of those three big three, Being, Mind and Form or as Ken Wilber wrote in its <u>Sex, Ecology and Spirituality the Spirit of Evolution</u> **[8]**

*With Kant, each of these spheres is differentiated and set free to develop its own potentials without violence...These three spheres, we have seen, refer in general to the dimensions of "it", of "we", and of "I"...In the realm of "itness" or empiric-scientific truths, we want to know if propositions more or less accurately match the facts as disclosed...In the realm of "I-ness", the criterion is sincerity...And in the realm of "we-ness" the criterion is goodness, or justness or relational care and concern...What is required, of course, is not a retreat to a predifferentiated state...what is required is the integration of the Big Three. And that, indeed, is what might be called the central problem of postmodernity...how does one integrate them?*

Science, philosophy and ontology and its integration implies a science that looks for truth but with an *openness* criterion, a philosophy that looks for *oneness* without the reductionism tendency but also an ontology that looks for *wholeness* as a fundamental principle.

To manage complexity properly it is very essential to have a basic structure and among the possible conceptual structures we can have
- a binary or dual one and
- a thirdness one that by mathematical inevitability becomes a sevenfold structure

This sevenfold structure has been used successfully in many fields in a natural way but also recently we can find its application in many other fields too. For example the three-"tiers" architectural approach to client/server solutions and which looks for separating the various components of a client/server system into three "tiers" of services that must come together to create an application is precisely a solution whose main aim is to manage the changing complexity and which requires a basic hierarchy that starts with the service to the client. A tool must be adapted in every case so that we can avoid what can be named the "Galileo syncretic whole reasoning" about that sevenfold structure. Maybe Galileo powerful mind "saw" the powerful explaining capacity of this structure but his time was just the beginning of a science in which the binary or dual structure would prevail.

Structure and environment are the main starting point of the new sciences of complexity or in the studies of complex adaptive systems in which adaptation to the environment implies always a minimum conceptual structural framework.

*User services* or interface or environment, *business services* or an adaptive plan, and *data services* or representation of changing structures in a general way, are just

**35**

some of the new applications we can found in the General Systems Sciences or Sciences of complexity of this sevenfold framework with which we definitely transcend the dualistic worldview and which is really very different from the holistic paradigm, and not just from the conceptual point of view but even more important from the practical point of view.